\documentclass{llncs}
\usepackage{makeidx}  

\usepackage{xspace}

\usepackage{graphicx}

\usepackage{todonotes}
\presetkeys{todonotes}{inline,color=blue!30}{}

\usepackage[T1]{fontenc}

\usepackage[inline]{enumitem}

\usepackage{hyperref}

\newcommand{\fnurl}[2]{\href{#2}{#1}\footnote{\url{#2}}}

\newcommand{\LoIDE}{\textbf{\textit{L}o\textsf{IDE}}\xspace}
\newcommand{\embasp}{\protect\textsc{embASP}\xspace}
\newcommand{\EmbASPServerExecutor}{\textbf{\textit{\embasp Server Executor}}\xspace}
\newcommand{\ace}{\textit{Ace}\xspace}

\begin{document}
\title{\LoIDE: a web-based IDE for Logic Programming\\Preliminary Technical Report}
\titlerunning{\LoIDE - Preliminary Technical Report}  
%
\author{Stefano Germano\inst{1} \and Francesco Calimeri\inst{1} \and Eliana Palermiti\inst{1}}
\authorrunning{Germano et al.} 
%
%
\institute{
	Department of Mathematics and Computer Science \\
	University of Calabria \\
	Via Bucci, Cubo 30B - 87036 Rende, Italy \\
	\email{\{germano,calimeri\}@mat.unical.it}
	\email{eliana.palermiti@gmail.com}
}

\maketitle              

\begin{abstract}
Logic-based paradigms are nowadays widely used in many different fields, also thank to the availability of robust tools and systems that allow the development of real-world and industrial applications.

In this work we present \LoIDE, an advanced and modular web-editor for logic-based languages that also integrates with state-of-the-art solvers.

\end{abstract}
\section{Introduction}\label{sec:introduction}

In the latest years, declarative paradigms and approaches to solving problems increasingly crossed the border of academia: it and has been going beyond theoretical studies in order to set into real world.
This is especially the case for logic-based formalisms; indeed, after years of theoretical results, the availability of solid and reliable systems supporting such formalisms made viable the implementation of effective logic-based solutions, even in the industrial context.

Along with such improvements in solver technology, the lack suitable engineering tools for developing programs started to be properly addressed; for instance, one might think of the work carried out by the Answer Set Programming (or AnsProlog) community~\cite{Baral:2003:KRR:582493,DBLP:journals/cacm/BrewkaET11}, that explicitly addressed issues like writing, debugging and testing Answer Set programs as well as embedding them into external, traditionally-developed systems.\cite{DBLP:journals/aim/ErdemGL16,DBLP:conf/rweb/LeoneR15}

At the same time, scenarios of computing significantly changed as well, now heavily relying on network connections and tools; in this context, the web-application paradigm, granting accessibility via a normal web-browser or even mobile devices, independently from the device in use, become very popular.
Indeed, many existing desktop applications have been ``ported'' to the web, and many new applications have been created specifically with this paradigm in mind.
Moreover, the JavaScript language has became a real cross-platform language, due to it's availability on all types of devices, ranging from servers to Internet of Things (IoT); it has been improved with many interesting features that made it an ideal language for developing not only small scripts, but also fully-fledged applications; and, fortunately, cloud-computing technologies significantly eased development, deploying and use of such applications.

In this scenario, many tools for software development have been released as web-applications, from simple text editors to integrated development environments (IDE).
As a result, code editors for many different programming languages are available to be used via a web browser, like \href{https://codeanywhere.com}{Codeanywhere}, \href{https://jsfiddle.net}{JSFiddle}, \href{https://c9.io}{Cloud9},  \href{http://codepad.org}{codepad}, \href{https://jsbin.com}{JS Bin}, \href{http://www.cssdesk.com}{CSSDesk}, \href{https://codepen.io}{CodePen}, \href{https://repl.it}{repl.it}, \href{https://codebunk.com}{CodeBunk}, and more.\\
Even developers of classic and famous development environments, like the Eclipse Foundation and Microsoft, have released cloud-based environments, respectively \href{https://eclipse.org/che}{Eclipse Che} (joint with Orion and Eclipse Dirigible) and \href{https://azure.microsoft.com/services/visual-studio-team-services}{Visual Studio Team Services} (formerly Visual Studio Online), that include also powerful IDEs.

Editors for logic programming are no exception: several have been build, from very simple playgrounds like the \href{https://repl.logicblox.com}{LogiQL REQPL} to more complex and complete editors like the \href{http://adams.cs.kuleuven.be/idp}{IDP Web-IDE}~\cite{DBLP:journals/corr/DassevilleJ15},  \href{http://swish.swi-prolog.org}{SWISH}~\cite{DBLP:journals/corr/WielemakerLR15} and the \href{http://editor.planning.domains}{PDDL Editor}~\cite{muise2016planning}.\\
In addition, specific editors for ASP programs like \href{http://potassco.sourceforge.net/clingo.html}{Clingo in the Browser}, \href{http://www.kr.tuwien.ac.at/research/systems/dlvhex/demo.php}{dlvhex Online Demo} and \href{http://asptut.gibbi.com}{Answer Set Programming for the Semantic Web - Tutorial} have been proposed; however, these are quite ``simplistic'', and at an early stage of development and, furthermore, it's worth noticing that each web-editor for logic programming that has been introduced to date is intended for a specific language, or even for a specific solver.\\
This raise some issues about interoperability, and limit the usage of these tools.

\medskip

In this paper we present \LoIDE, a web-based IDE for Logic Programming that explicitly addresses interoperability and flexibility, supporting multiple formalisms and corresponding solvers.

The remainder of the paper is structured as described next.
In Section~\ref{sec:loide} we describe the \LoIDE project and its main features and then we explain some implementation details of the different components of the project (Section~\ref{sec:implementation}).
The paper concludes with a comparison with similar projects (Section~\ref{sec:other-ide}) and a discussion about the future developments of the \LoIDE project.

\section{The \LoIDE project}\label{sec:loide}

\begin{figure}
	\includegraphics[width=\textwidth]{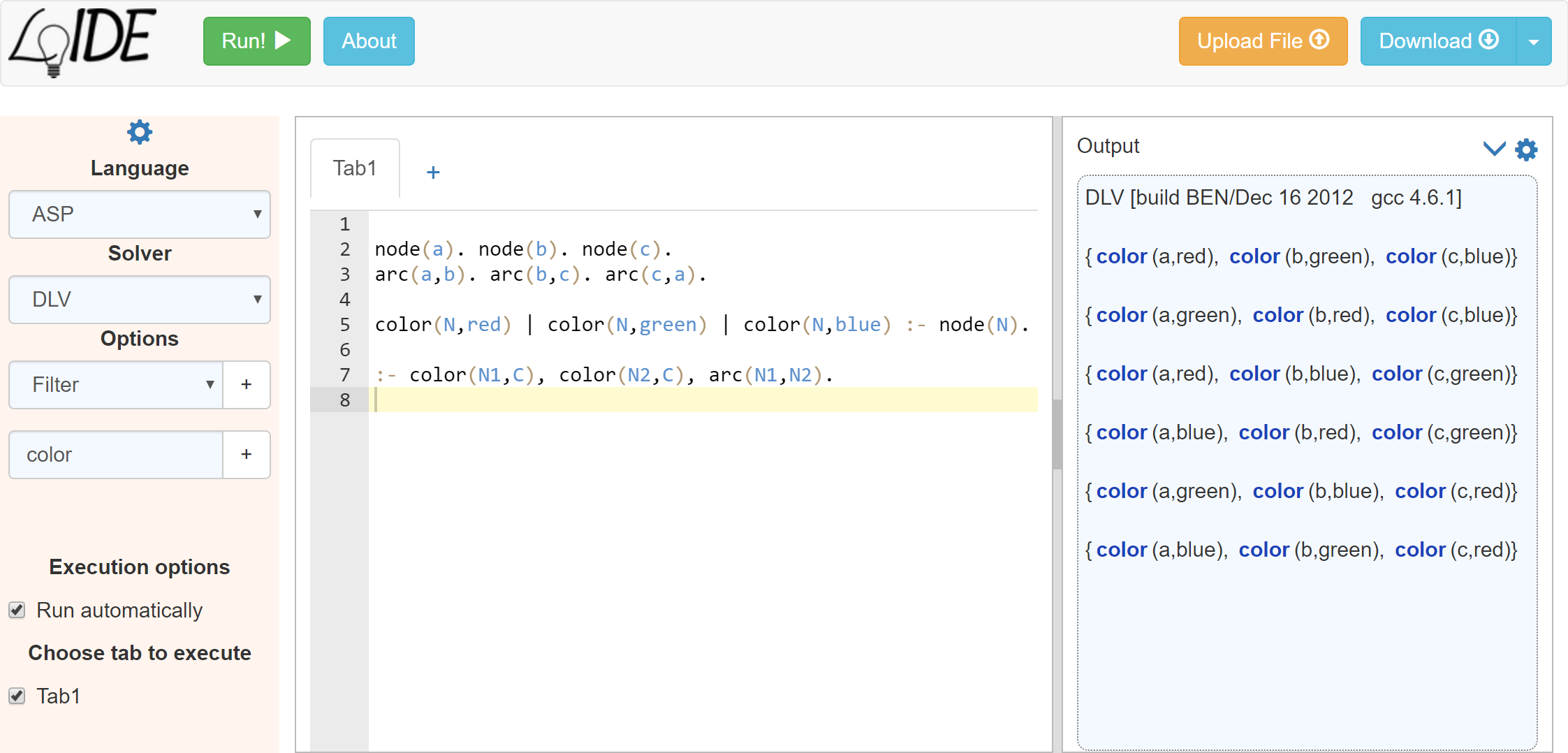}
	\caption{An ASP program addressing a toy instance of the 3-colorability problem and the corresponding run performed by the DLV system via \LoIDE.}
	\label{fig:screenshot_3-col}
\end{figure}

The main goal of the \LoIDE project is the release of a modular and extensible web-IDE for Logic Programming using modern technologies and languages.
The \LoIDE IDE will provide advanced features specifically tailored for Logic Programming; it has been conceived in order to be extended during time, and will include as many logic-based languages and solvers as possible.
A further goal of the project is to provide a web-service with a common set of APIs for different logic-based languages; at the time of writing, this is still at an early stage of development.

\LoIDE is provided as open-source software (OSS) and it's publicly available at \url{https://goo.gl/sDGMhA}.
Moreover, we released it as Free Software\footnote{Under MIT License. \url{https://goo.gl/nrXtN4}}, with the explicit aim of helping the community of researchers and developers, that are free to study, use, distribute and even improve the project.\\
A prototypical running demo is available at \url{https://goo.gl/s4g6zA}.

\subsection{Features of the IDE}\label{subsec:features}

The \LoIDE IDE provides all the basic features of text and code editing that can be of use for Logic Programming.
We started from basic features available in \fnurl{\ace}{https://ace.c9.io}, a JavaScript embeddable code editor that constituted the base for \LoIDE (see Section~\ref{subsec:GUI}).
Among them the most relevant, we mention here:
\begin{description}
	\item[\emph{indentation}:] automatically indent and outdent the code;
	\item[\emph{document size}:] handles huge documents;
	\item[\emph{key bindings}:] has fully customizable key bindings (including vim and Emacs modes);
	\item[\emph{search and replace}:] ability to search and replace inside the text using regular expressions;
	\item[\emph{brace matching}:] highlight matching parentheses;
	\item[\emph{mouse gestures}:] drag and drop text using the mouse;
	\item[\emph{advanced cursors management}:] multiple cursors and selections;
	\item[\emph{clipboard management}:] cut, copy, and paste functionality;
	\item[\emph{themes}:] over 20 themes available (TextMate/Sublime Text .tmtheme files can be imported).
\end{description}

We extended such basic functionalities in order to properly meet the specific requirements of Logic Programs development.

\paragraph{Syntax highlighting.}
\ace supports syntax highlighting, and it already covers 110 languages; unfortunately, it did not support the logic-based languages we were interested in.
Relying on the specifications for cross-browser syntax highlighting, we introduced a basic support for syntax highlighting of ASP programs.
We plan to include other languages as soon as the support for their specific solvers will be added to \LoIDE.

\paragraph{Editor layout and appearance.}
The user can customize layout and appearance of the ``Input'' and the ``Output'' fields of the IDE.
The user can change theme and fonts of each part of the interface, independently.
Moreover, size and position of the two fields are customizable as well.
It's worth noticing that the all the options are automatically saved in the \fnurl{Web Storage}{http://www.w3.org/TR/webstorage}, in order to make the persistent also across different the current browser user sessions.

\paragraph{Output highlighting.}
One of the most annoying aspect of developing and testing logic programs in practice is the need for checking output in test cases: since output is often constituted of a (possibly long) list of instances of many predicates, it can be quite tricky.
Most ASP solvers allow to filter predicates, but this does not solve the problem and it is not a very flexible solution.
\LoIDE features an ad-hoc output highlighting: when the user selects an element of the output (for instance, a predicate name), all the elements with the same ``name'' will be automatically highlighted.
The user can dynamically play with such highlighting, and as a consequence the analysis of the results might dramatically simplified.\footnote{More features for improving comprehension of the results will be added, such as different forms of visualization}.

\paragraph{Keyboard shortcuts.}
\LoIDE supports several many keyboard shortcuts.
We properly extended the typical code-editors shortcuts provided by \ace\footnote{\url{https://github.com/ajaxorg/ace/wiki/Default-Keyboard-Shortcuts}} for
\begin{enumerate*}[label={\alph*)}]
	\item Line operations,
	\item Selection,
	\item Multi-cursors,
	\item Go-to,
	\item Find/Replace,
	\item Undo/Redo,
	\item Indentation,
	\item Comments and
	\item Word/Character variations;
\end{enumerate*}
adding specific shortcuts to Save and Load files and to Run the logic program.

\vspace{1em}

Furthermore, we implemented other custom features around the \ace-based stem.

\paragraph{Multiple file support.}
When dealing with real-world problems in practice, logic programs often results to be splitted in more than one file (think, for instance, of the obvious separation between problem specification and problem instances).
\LoIDE explicitly support multiple files management: the user can create and manage many different tabs, and also selectively decide which one has to be composed into the actual program to run.

\paragraph{Options.}
Settings of \LoIDE can be customized, along with the behaviour of the underlying systems of use.
The user can select the logic programming language of choice; for each language, the solver to be used in order to run the program can be set as well.
Moreover, specific options can be selected for each solver, with predefined typical settings available for the most common.
There are also more general options; for instance, the user can ask to automatically run the program at the end of each statement, so that the output dynamically changes as the user is crafting the program; this increase the interaction with the system. Furthermore, it might significantly ease the development of non-trivial programs, and be of great help in educational settings (think about a Logic Programming class, for instance).

\paragraph{Import and Export files.}
Content of the editor, all options and outputs can be downloaded locally to the device of use as \emph{JSON} files, and later restored, possibly over a different device (also by means of Drag-and-Drop, if the device supports it).
Such feature is crucial for practically provide the user with a working environment which is virtually immaterial and free from specific physical work stations.
Of course, also logic program being edited can be saved.

\section{Implementation}\label{sec:implementation}
We provide next some insights on the design and implementation of \LoIDE.

\subsection{General Architecture}

\begin{figure}
	\includegraphics[width=\textwidth]{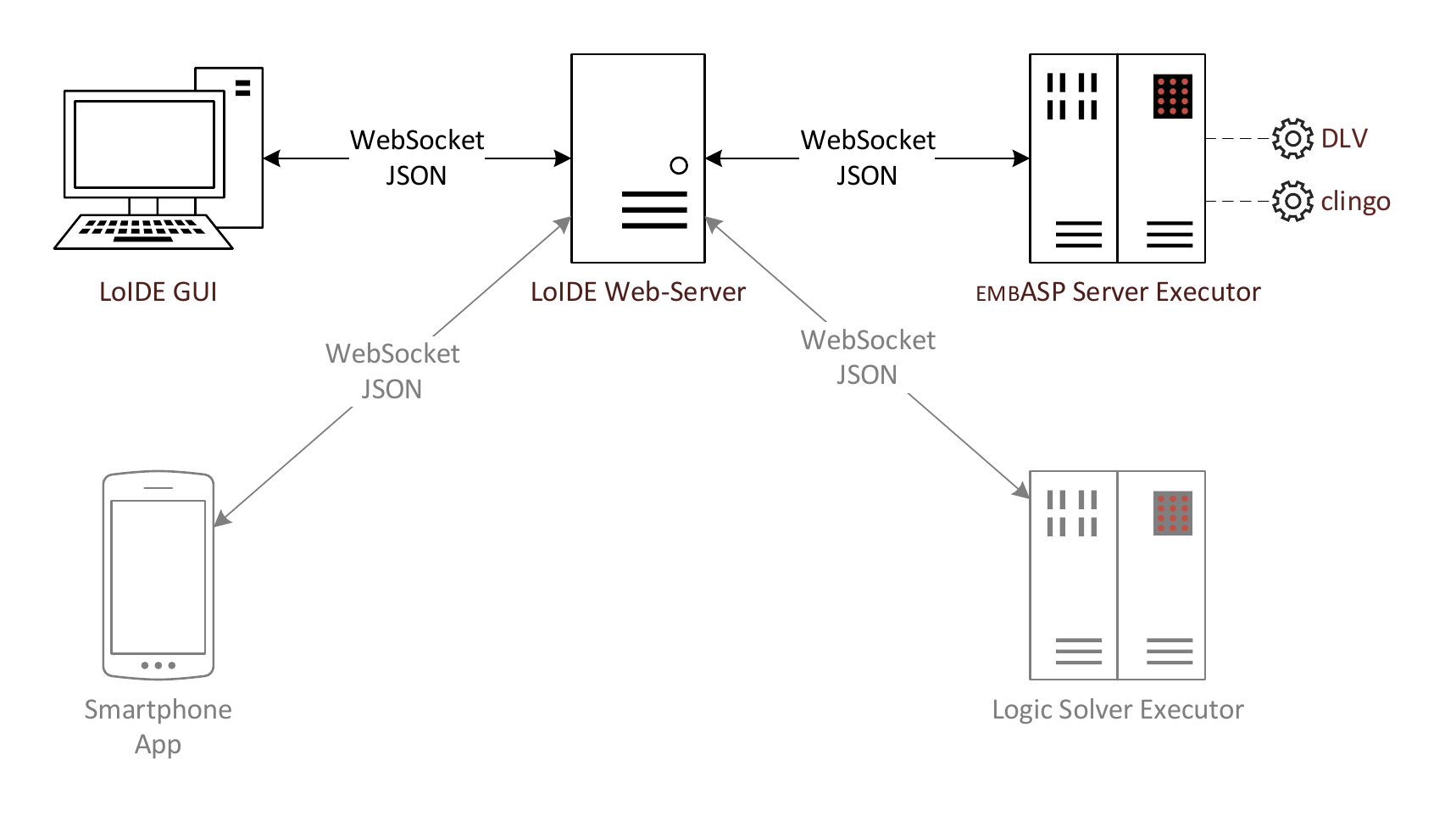}
	\caption{Architecture of the \LoIDE project}
	\label{fig:architecture}
\end{figure}

The system architecture, relying on a typical client-server framework, is draft in Figure~\ref{fig:architecture}.
The back-end (or \emph{server-side} component) consists of the main \LoIDE Web-Server, developed using \fnurl{\emph{Node.js\textsuperscript{\textregistered}}}{https://nodejs.org}, which exposes \emph{API} that can be used by the client.

The front-end (or \emph{client-side} component) consists of the Graphical User Interface (GUI) of the IDE, developed using modern web technologies such as \emph{HTML5}, \emph{CSS3} and the \emph{JavaScript} programming language.

The execution of the logic solvers is not performed directly from within the main \LoIDE web-server; a dedicated component, the \EmbASPServerExecutor, is in charge of this, instead.
This choice is due to the aim of keeping the system modular and extensible; indeed, such  modularity allows to easily support different ``executors'' and ease the management of additions, upgrades, and security issues.

All the components communicate using the \fnurl{\emph{WebSocket}}{https://tools.ietf.org/html/rfc6455} communication protocol and the \fnurl{\emph{JSON}}{http://www.json.org} data-interchange format.

Thanks to the structure of the project and the use of standard and common technologies between all the components, modules can be added or modified in a straightforward way while maintaining the scalability of the whole architecture as can be seen in the lower part of Figure~\ref{fig:architecture}.

\subsection{Back-end -- The \LoIDE Web-Server}

The \LoIDE Web-Server has been developed using Node.js\textsuperscript{\textregistered}.
In order to effectively use \emph{WebSockets}, we relied on the \fnurl{\texttt{socket.io}}{https://socket.io} package to enables real-time bidirectional event-based communication between the client and the server; the package provided us with means for enabling several useful features, such as Reliability, Auto-reconnection support and Disconnection detection.

\paragraph{\LoIDE APIs.} \
As mentioned before, one of the aims of the \LoIDE project is to develop a set of (Web) APIs for easily and efficiently controlling different solvers over different logic programming languages, using the WebSocket protocol and the JSON format.
Specifications and implementation are at the first stage; currently, a call type is available, that given the description of the \texttt{language}, the \texttt{solver}, the list of \texttt{options} and the \texttt{program}, executes the solver over the program and returns either the \texttt{output} of the solver or any \texttt{error} messages.
See \href{https://goo.gl/6XJeDN}{\LoIDE API documentation}\footnote{\LoIDE APIs. \url{https://goo.gl/6XJeDN}} for further details.

\subsection{Front-end -- The \LoIDE GUI}\label{subsec:GUI}

The front-end uses modern standard web technologies (HTML5, CSS3, JavaScript); hence, \LoIDE is compatible with virtually any device currently available.
We used some popular frameworks and libraries with the aim of improving user experience and making the IDE robust and powerful.
\begin{description}
\item[\ace] \fnurl{\ace}{https://ace.c9.io} is a JavaScript embeddable code editor. It matches the features and performance of native editors such as Sublime, Vim and TextMate and it can be easily embedded in any web page and JavaScript application. \ace is maintained as the primary editor for \href{https://c9.io}{Cloud9 IDE} and is the successor of the Mozilla Skywriter (Bespin) project. \ace is a community project and its source code is hosted on GitHub and released under the BSD license.
\item[Bootstrap] \fnurl{Bootstrap}{https://getbootstrap.com} is the most popular front-end component library framework for developing responsive, mobile-first projects on the web. Bootstrap is an open source toolkit for developing with HTML, CSS, and JavaScript and its source code is hosted on GitHub and released under the MIT license.
\item[jQuery and its UI Layout plugin] \fnurl{jQuery}{https://jquery.com} is a fast, small, and feature-rich JavaScript library. It makes things like HTML document traversal and manipulation, event handling, animation, and Ajax much simpler with an easy-to-use API that works across a multitude of browsers. With a combination of versatility and extensibility, jQuery has changed the way that millions of people write JavaScript. \\
The \fnurl{jQuery UI Layout plugin}{http://plugins.jquery.com/layout} allows to create advanced UI layouts with sizable, collapsible, nested panels and tons of options. It integrates with and enhances other UI widgets, like tabs, accordions and dialogs, to create rich interfaces.
\item[bimap] \fnurl{BiMap}{https://github.com/alethes/bimap} is a powerful, flexible and efficient JavaScript bidirectional map implementation. Enables fast insertion, search and retrieval of various kinds of data. A BiMap is like a two-sided JavaScript object with equally immediate access to both the keys and the values.
\item[keymaster.js] \fnurl{Keymaster}{https://github.com/madrobby/keymaster} is a simple micro-library for defining and dispatching keyboard shortcuts in web applications.
\end{description}

\begin{figure}
	\includegraphics[width=\textwidth]{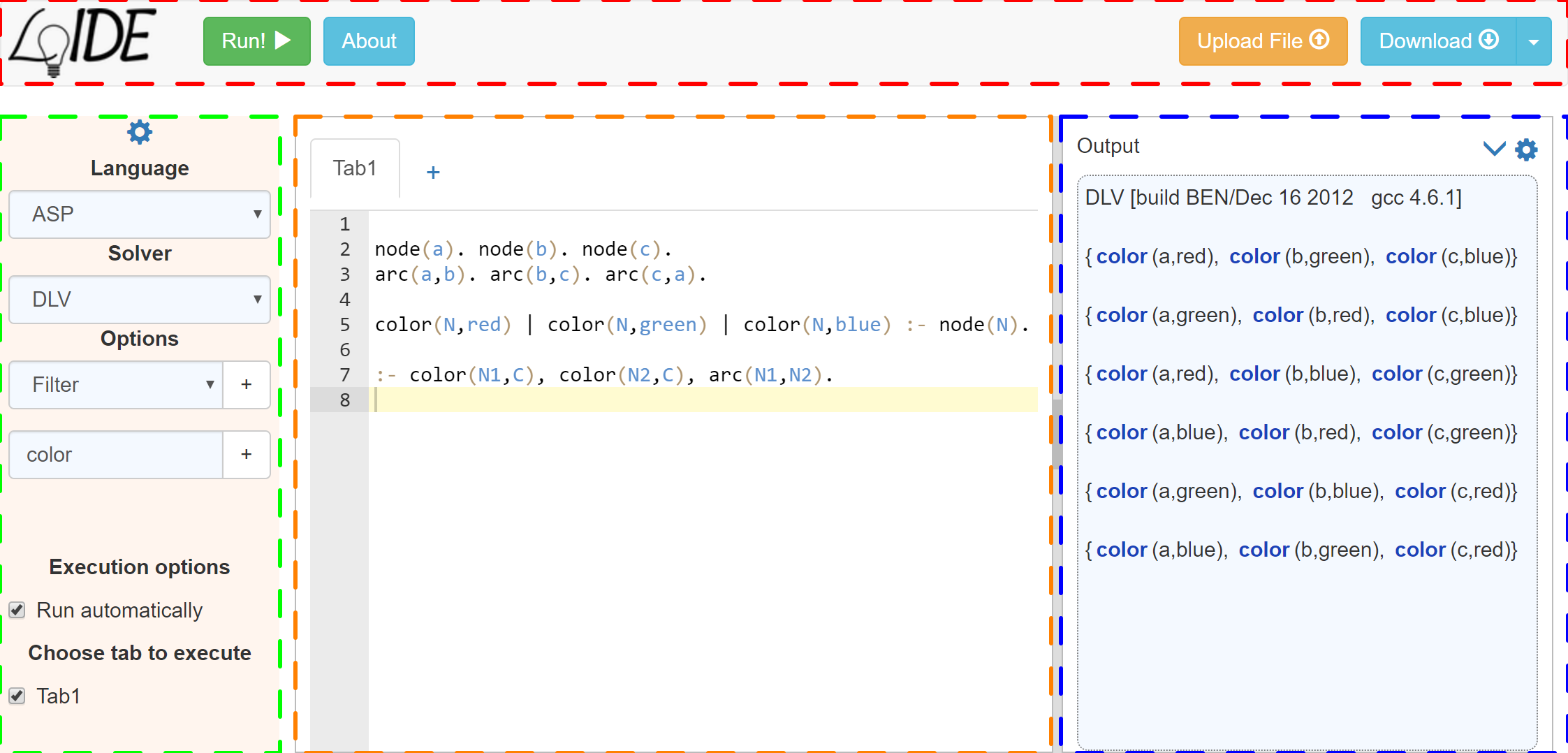}
	\caption{The main components of the web-based GUI.}
	\label{fig:screenshot_3-col_wb}
\end{figure}

The web-based GUI is divided into $4$ different parts (Figure~\ref{fig:screenshot_3-col_wb}).

At the top the \emph{navigation bar} is shown, highlighted in red in Figure~\ref{fig:screenshot_3-col_wb}, it contains proper \emph{Run}, \emph{Upload} and \emph{Download} buttons.
In the middle of the interface, highlighted in orange in Figure~\ref{fig:screenshot_3-col_wb}, the \emph{code editor} contains the editing tabs holding the program(s) to execute.
At the right side, highlighted in blue in Figure~\ref{fig:screenshot_3-col_wb}, the \emph{output panel} dynamically shows the output of the computations and the link to the editor's layout options.
At the left side, highlighted in green in Figure~\ref{fig:screenshot_3-col_wb}, the \emph{IDE options panel} contains all the options described in the Section~\ref{subsec:features}.
This panel can be automatically toggled in order to save space for the main editor.

The layout is built using the Responsive Web Design (RWD) approach, i.e., it automatically adapts to the viewing environment and offers the possibility to be viewed on different devices with almost the same User Experience.

\subsection{The \EmbASPServerExecutor}\label{subsec:ese}
In order to decouple the management of the web-requests from the execution of the logic programming solvers, we developed \EmbASPServerExecutor as a completely different application; it is even implemented in a different programming language.

\EmbASPServerExecutor is a Java server application that is able to execute ASP programs with different solvers.
It has a the usual structure of a Java web-app with the following modules:
\begin{enumerate*}[label=(\roman*)]
	\item Control,
	\item Model,
	\item Service, and
	\item Resources.
\end{enumerate*}
For space reasons, we do not discuss them in details, as the names are self-explanatory.
\EmbASPServerExecutor runs on top of Apache Tomcat\textsuperscript{\textregistered} and it exposes a set of APIs that can be used to invoke the solvers.
In order to execute the desired solver, it makes use of \href{https://goo.gl/mE5kAo}{\embasp}\footnote{\embasp. \url{https://goo.gl/mE5kAo}}~\cite{DBLP:conf/ppdp/FuscaGZACP16}.
\embasp is a framework for the integration (embedding) of Logic Programming in external systems for generic applications; it helps developers at designing and implementing complex reasoning tasks by means of solvers on different platforms.

Similarly to \LoIDE, \EmbASPServerExecutor is provided as open-source software (OSS) and it's publicly available at \url{https://goo.gl/2WUeb4}.
Moreover it's likewise released as Free Software\footnote{Under MIT License. \url{https://goo.gl/VfPknG}}.

\section{Related Works}\label{sec:other-ide}

The work herein presented is naturally comparable to other Logic Programming IDEs and other web-based editors.

Several stand-alone, ``native'' editors and IDEs have been proposed for Logic Programming over different platforms; we refer the reader to the ample   literature on the topic~\cite{online:iGROM,Sureshkumar07ape:an,DBLP:journals/logcom/RiccaGSDGL09,wielemaker:2011:tplp,DBLP:conf/lpnmr/FebbraroRR11,DBLP:conf/lpnmr/OetschPSTZ11,DBLP:journals/tplp/BusoniuOPST13,DBLP:conf/ki/StrobelK14};  moreover, many web-based editors have been recently introduced~\cite{DBLP:conf/ki/StrobelK14,DBLP:journals/corr/WielemakerLR15,DBLP:journals/corr/DassevilleJ15,muise2016planning,DBLP:journals/corr/MarcopoulosRZ17}.

All tools and environments share the same core of basic features, many of them are quite stable, some are already well-known.
\LoIDE, similarly to most of the web-based editors, currently has less features with respect to the ``native'' ones; however, even if quite young, is already stable and allows to effectively take advantage from logic programming without the need for installing or downloading anything from almost any platform connected to the Internet; furthermore, it could even run locally on any device featuring the Node.js runtime, with a few additional configuration steps.
Some tools (as SWISH, for instance) rely on platforms that provides functionalities via specific APIs over HTTP; it is worth noting that this is not the case of \LoIDE. 
Indeed, it started from Answer Set Programming, where no such platforms were available; it makes use of the \EmbASPServerExecutor, implemented on purpose, that makes the project also more general and extensible.

All mentioned editors have peculiar, sometimes very interesting features; however, each one is tailored to a specific language and tightly coupled with some specific solver(s) in the back-end.
On the other hand, the aim of \LoIDE is to have a robust platform that seamlessly integrates different languages and different solvers.
We do believe that this approach is more general, and could foster the use of logic programming in many contexts, especially in practical context and in education, also fruitfully promoting exchanges among the various communities in the logic programming area.

\section{Conclusion and Future work}\label{sec:conclusion}
In this paper we presented the \LoIDE project, an advanced and modular web-editor for logic-based languages that is also capable of integrating with state-of-the-art solvers.
Even if already equipped with relevant features that make it effectively usable in practice, the project is still at an early stage of development; hence, we have already identified many future works and improvements.

We planned to extend editing capabilities by
\begin{enumerate*}[label={\alph*)}]
	\item auto-complete/intellisense,
	\item language-based syntax checking,
	\item snippets and linting support,
	\item plugins/extensions support,
	\item sharing of the programs in the cloud (for instance supporting Dropbox and Gist API).
\end{enumerate*}
We are also planning to explicitly support the check for the ``type'' of file for each tab; indeed, it's a common practice in some logic languages to have different definitions in different files (think about the ``domain text'' and the ``problem text'' of \emph{PDDL}).

Currently, \LoIDE sends the results back to the client when the solver finishes the job, given that the framework used by the \EmbASPServerExecutor (Section~\ref{subsec:ese}) does not handle ``output streams''; however, the \LoIDE Web-Server easily allows to serve back the results as soon as they are produced by the solver.
An important improvement will allow such mechanism in order to save bandwidth and speed up the reception of the results from the solvers.

We plan to improve the deployment using container technologies (for instance Docker) and cloud-computing services (like Amazon Web Services, Microsoft Azure or Google Cloud Platform) in order to make it easier and more robust.

Advanced features like visualization techniques, such as~\cite{DBLP:conf/lpnmr/AmbrozCJWW13,DBLP:conf/iclp/CliffeVBP08,DBLP:journals/corr/LapauwDD15,DBLP:conf/inap/KloimullnerOPT11}, will also further simplify output comprehension, and the possibility to save files in the user account or over cloud services will allow also pave the way to team work and collaborative editing.
We are working to support more \emph{executors} (web-services), logic-based \emph{languages} and \emph{solvers} (engines), in order to increase the audience of the project; the addition of an interactive tutorial could also allow the users to became more familiar \LoIDE and with declarative programming.

%
%
\bibliographystyle{splncs03}
\bibliography{references}

\end{document}